# Effect of Succinonitrile on Ion Transport in PEO-based Lithium-Ion Battery Electrolytes

Sipra Mohapatra[1,†], Shubham Sharma[1,†], Aman Sriperumbuduru[1], Srinivasa Rao Varanasi[2], and Santosh Mogurampelly[1,*]

[1]Department of Physics, Indian Institute of Technology Jodhpur, Karwar, Rajasthan 342027, India
[2]Department of Physics, Sultan Qaboos University, Al-Khoud 123, Muscat, Oman.

**ABSTRACT:**

We report the ion transport mechanisms in succinonitrile (SN) loaded solid polymer electrolytes containing polyethylene oxide (PEO) and dissolved lithium bis(trifluoromethane)sulphonamide (LiTFSI) salt using molecular dynamics simulations. We investigated the effect of temperature and loading of SN on ion transport and relaxation phenomenon in PEO-LiTFSI electrolytes. It is observed that SN increases the ionic diffusivities in PEO-based solid polymer electrolytes and makes them suitable for battery applications. Interestingly, the diffusion coefficient of TFSI ions is an order of magnitude higher than the diffusion coefficient of lithium ions across the range of temperatures and loadings integrated. By analyzing different relaxation timescales and examining the underlying transport mechanisms in SN-loaded systems, we find that the diffusivity of TFSI ions correlates excellently with the Li-TFSI ion-pair relaxation timescales. In contrast, our simulations predict distinct transport mechanisms for Li-ions in SN-loaded PEO-LiTFSI electrolytes. Explicitly, the diffusivity of lithium ions cannot be uniquely determined by the ion-pair relaxation timescales but additionally depends on the polymer segmental dynamics. On the other hand, the SN loading induced diffusion coefficient at a given temperature does not correlate with either the ion-pair relaxation timescales or the polymer segmental relaxation timescales.

*Address for correspondence: santosh@iitj.ac.in
[†]Equal first author contribution

# 1. INTRODUCTION

The growing demand for new generation portable electronic devices is driving the technological development of rechargeable solid-state batteries. The Li-ion batteries offer many exciting properties such as high energy density, lightweight, flexibility, and longevity, among many other properties.[1–3] Solid polymer electrolytes (SPEs) are becoming increasingly attractive electrolyte materials for lithium-ion batteries because of their safety, flexibility, and mechanical strength.[4–6] They also have good electrochemical stability, low flammability, and toxicity with the ability to form suitable interfacial materials with the electrodes, thereby eliminating the need for a separator. Polyethylene oxide (PEO) is a promising candidate for SPE applications in Li-ion batteries. However, they possess a very low ionic conductivity (in a range of $10^{-8}$ to $10^{-4}$ S/cm at room temperature) compared to liquid electrolytes, making them less practical for commercial applications.[7,8] Therefore, there is a quest for new generation electrolyte designs offering increased ionic conductivity of PEO-based SPEs. In this context, a thorough understanding of the fundamental transport mechanisms is necessary to provide design strategies for new generation electrolytes.

The PEO is a semi-crystalline polymer with a chemical structure of $H-(O-CH_2-CH_2)_n-OH$. The ether oxygens (EO) units in PEO act as a promising association site for Li-ions, making the PEO a good ion-solvating candidate for promoting faster ion transport for battery applications. A study by Fenton et al. in 1973 found that alkali metals get easily dissolved in PEO to form conductive complexes.[9] In 1975, Peter V. Wright, a polymer chemist from Sheffield, first found PEO as a host for sodium and potassium salts for solid-based electrical conductor polymer-salt complex.[10,11] A mechanistic study by Michel Armand established in 1983 that the ionic motion is due to the amorphous region in polymer-salt complexes.[12] Soon after that, explaining the physics of PEO became an active research area to understand its ion transport mechanism, chemical and physical properties at the electrode/electrolyte interface.[2]

The ion transport mechanism of neat amorphous PEO melt has been investigated previously using both experimentally and theoretically.[13–16] It was reported that the amorphous phase of PEO activates the chain segments, which aids the ion transport, whereas its crystalline phase slows down the polymer chain dynamics.[14] In SPEs, the Li-ions are located at specific coordinate sites near the EO groups in polyethylene oxide chains, which undergo a constant segmental motion. As a result, the lithium-ion hops from one EO site to the other along the backbone of the polymeric chain and intermittently jumps from one chain to another under the motion of segmental motion of polymeric chains. Borodin and Smith studied the ion transport mechanism in amorphous PEO/LiTFSI, a PEO-based SPEs. They observed that the diffusion of Li and TFSI ions are strongly coupled to PEO ether oxygen atom displacement and its conformational dynamics.[14]

More importantly, traditional lithium salts (LiClO$_4$, LiPF$_6$, LiAsF$_6$, LiBF$_4$, etc.) aggregate strongly in SPEs[17] Ion aggregation, which depends on the size of anion, increases with salt concentration and temperature due to the delocalization of charge on large-sized anions.[18] The bis(trifluoromethane)sulphonamide anion (TFSI$^-$) class of ILs is popular for Li-ion battery applications because of its large size. Piotr et al. investigated the correlations in ion transport in LiTFSI electrolytes.[19] They found that irrespective of the type of cation, the motion of cation and anion (TFSI) are strongly correlated, resulting in reduced ionic conductivity.[19] In 2000, Borodin and Smith studied the dynamics of cation and anion in LiI-doped diglyme and poly(ethylene oxide) solutions having (EO: Li)=15:1 and 5:1[13] and found that ion-aggregation is more significant in the case of diglyme/LiI solution which results in a decrease of ion mobility and conductivity. Further, they also observed that EO-Li association lifetime is highly correlated with torsional autocorrelation time for -O-C-C-O- dihedrals.

To alleviate the problem of high crystallinity in neat SPEs, some solvents or plasticizers are typically added to the electrolyte increasing the amorphous nature of the polymer matrix.[20] Many experimentalists have studied the effect of adding different solvents to the PEO matrix on ionic conductivity. For instance, Ahn et al. incorporated 1-butyl-3-methylimidazolium-bis (trifluoromethane sulfonyl) imide (BMITFSI) in the PEO-LiTFSI complex, found that with the increase in BMITFSI content, the ionic conductivity increases, and the reported value is 0.32 mS/cm at room temperature.[21] Several molecular dynamics studies reported that the PEO matrix incorporated with different solvents shows better ionic conductivity and polymer flexibility.[13,22–27]

Due to plasticity and high polarity, succinonitrile (SN) is considered a potential additive in polymer electrolytes.[1,28–31] Experimentalists have observed a high ionic conductivity (~ 0.35 mS/cm at 30°C) and high mechanical stability because of decreased tensile strength in SPEs by appropriately optimizing the content of SN and PEO.[28] It is demonstrated that different polymer electrolytes P(VDF-HFP)–LiTFSI and P(VDF-HFP)–LiBETI show excellent mechanical properties when SN is used as a dispersant.[31] Gianni et al. have investigated the dynamical properties of succinonitrile by calculating the time dependency of single-molecule autocorrelation functions and found three distinct regimes of orientation relaxation.[1] Also, it is observed that the complex formed by PEO and alkali metal salts had shown fast alkali metal transport at 100°C.[9] While several experimental studies have been carried out to understand the influence of SN on ionic conductivity in a polymer matrix, a molecular-level understanding of the mechanisms of ion transport is not understood.[28–31] Particularly, a fundamental understanding of the factors influencing ion transport in SN-loaded systems is lacking. Molecular dynamics with the atomistic level resolution is a powerful tool to gain deeper physical insight into the ion-transport mechanisms. Therefore, we considered the atomistic molecular dynamics simulations comprising of PEO-LiTFSI matrix dispersed with succinonitrile (SN) molecules at the desired loading. Our primary motivation for this study is: (i) To examine the diffusivities of ions in SN loaded PEO-LiTFSI (ii) To understand the governing factor of ionic diffusivities through the ion-pair interactions and polymer segmental motion, and finally, (iii) to explain the transport mechanisms of ions inside the PEO matrix solvated with Li-TFSI and SN molecules. Ion-pair relaxation times and dihedral relaxation times were calculated to understand the interplay of interactions of Li-ion with the counterions and polymer.[13,14,16,23,32–36]

Towards the above-outlined objectives, we analyzed the MD simulations of SN-PEO-LiTFSI by varying temperature and the weight percentage of SN molecules. The effect of succinonitrile on ion transport mechanisms was investigated using fully atomistic simulations of 200 ns and longer, depending on the loading and temperature. We have calculated the diffusion coefficients of both $Li^+$ (cation) and $TFSI^-$ (anion) at a concentration (EO: Li = 12.5:1) and different weight percentages of SN in PEO-LiTFSI melt. The temperature effects on ion transport were studied by performing simulations at different temperatures ranging between 375 K to 575 K. Together, our results indicate that the SN-induced changes in ion-pair relaxations and polymer dynamics arising from ion-ion interactions and ion-polymer interactions, respectively, are the main mechanisms of ion transport in SN-loaded PEO-LiTFSI electrolytes.

The organization of the rest of the paper is as follows: In section **2,** we present the simulation details, including the interaction potential field and force field parameters used, initial system setup, equilibration, and production protocols. Section 3.1 presents the results diffusion coefficients calculated from mean-squared displacement (MSD) of Li and TFSI ions, respectively. In section **3.2,** we identify different mechanisms influencing ion mobility. Explicitly, in section **3.2.1,** we present the results for ion-pair autocorrelation functions, corresponding relaxation times with diffusivities, and discuss the correlations with ionic mobility. Section 3.2.2 presents the results of dihedral autocorrelation functions and corresponding relaxation times and examines their connection with polymer segmental motion. Finally, in section **4,** we summarized our results and concluded essential findings of understanding the ion transport mechanisms in SN-PEO-LiTFSI electrolytes.

## 2. SIMULATION METHODS

## 2.1. Interaction Potential and Force Fields.

We used atomistic molecular dynamics (MD) simulations to investigate the PEO-LiTFSI electrolytes loaded with SN molecules. The molecular dynamics simulations of the SN-PEO-LiTFSI system were performed using GROMACS 2020.2[37] (GROningen MAchine for Chemical Simulations) package employing the following general form of the interaction potential between different atoms:

$$U(r) = U^{bonded}(r) + \sum 4\epsilon_0 \left[\left(\frac{\sigma}{r_{ij}}\right)^{12} - \left(\frac{\sigma}{r_{ij}}\right)^6\right] + \sum \frac{q_1 q_2}{4\pi\epsilon_0 r_{ij}} \quad\text{(1)}$$

where $U^{bonded}$ describes the interactions arising from bonds, angles, and torsions in PEO. The bonded interactions in PEO were modeled with a harmonic potential $\frac{1}{2}k_r(r-r_0)^2$, the angles with $\frac{1}{2}k_\theta(\theta-\theta_0)^2$, and torsions with $\frac{1}{2}\sum_{n=1}^{4} C_n[1+(-1)^{n+1}\cos(n\phi)]$. The non-bonded interactions are modeled with the Lennard-Jones potential and the Coulombic electrostatic potential. The parameters in eq 1 are obtained from the set of Optimized Potentials for Liquid Simulations (OPLS) forcefield.[38,39] A scaling factor of 0.5 was used to calculate the effective nonbonded interactions associated with intramolecular atomic pairs separated by three bonds, and complete nonbonded interactions are considered for intramolecular atomic pairs beyond three bonds. The nonbonded parameters for cross-interaction terms between different atomic types were calculated using the geometric mixing rules.

The Lennard-Jones models with full charges on ionic species produce results that are inconsistent with experiments. For example, the mobility of ions decreases due to the strong ion-ion interactions by almost an order of magnitude, and the densities increases. The polarized charge models are better suited for computational approaches to reproduce the experimental densities and diffusion coefficient in comparison with experiments. However, they are computationally expensive. To overcome this, scaling of partial atomic charges is considered one of the judicious choices which produce better transport properties and densities with the less computational cost involved. Several authors used charge scaling to study the transport properties of electrolyte solutions.[40–46] Therefore, the total charge on ionic species was scaled to 0.8e to indirectly mimic the induced polarization effects in a mean-field like manner. Such an approach was previously shown to produce results compared to polarizable models and experiments.[34,47–49] The charge scaling approach is being routinely employed in atomistic MD simulations as an alternate approach to reproduce the dynamics of ionic species in comparison with experiments.[32,34,41,42,44,46,50–53] We obtained the diffusion coefficients of $0.1 \times 10^{-7}$ and $0.8 \times 10^{-7}$ cm$^2$/s and lithium and TFSI ions, respectively at 375 K for pure PEO-LiTFSI electrolyte and a density of 1.294 g/cm$^3$. Our charge scaling approach resulted in diffusion coefficient and density results that are comparable to the literature.[46,53–59] The partial charges of different atoms in our simulations are shown in Table **1**. The molecular structure with naming convention is provided in Figure **S1** in the SI.

**Table 1**. Partial charges on atom types used in our MD simulations

| Atom (PEO) | q (e) | Atom (TFSI) | q (e) | Atom (SN) | q (e) |
| --- | --- | --- | --- | --- | --- |
| O | -0.40 | F1 | -0.128 | C00 | 0.116 |
| C | 0.14 | CBT | 0.280 | C01 | -0.070 |
| H | 0.03 | SBT | 0.816 | C02 | -0.070 |
| C1 | 0.11 | OBT | -0.424 | H03 | 0.144 |
| Li | 0.800 | NBT | -0.528 | H04 | 0.144 |
| | | | | C05 | 0.116 |
| | | | | H06 | 0.144 |
| | | | | H07 | 0.144 |
| | | | | N08 | -0.335 |
| | | | | N09 | -0.335 |

**2.2. Initial System Setup.** The initial conformation of PEO was constructed as follows: We built a single polymeric chain of 50 repeating monomeric units with chemical structure H-[CH$_2$-O-CH$_2$]$_{50}$-H corresponding to a molecular weight of 2.2 kDa. Minimization was performed on this single PEO chain to get the desired bond lengths, angles, and torsion angles defined according to OPLS-AA force field parameters.[38,39] The minimized configuration of a single PEO chain was then inserted randomly in a simulation box to generate a less dense system consisting of 100 chains of PEO using Packmol software.[60,61] Then, the pure 100 chains of the PEO system were solvated with appropriate numbers of Li$^+$ cations and TFSI$^-$ anions corresponding to a salt concentration of EO: Li=12.5:1 by placing the ions randomly into the simulation box.

To build the complete system, the succinonitrile (SN) plasticizers were then dispersed randomly in bulk PEO-LiTFSI matrix at different loadings of 0, 2.2, 10, 15, 20 wt%s. The chemical structure of SN is [CN-CH$_2$-CH$_2$-CN] and has a molecular weight of 80.09 g/mol. It is a highly flexible solid plasticizer with a plastic crystalline organic molecular structure, stable between 233 and 331 K.[1] The system's density in the initial stage was chosen as low as needed to easily solvate the Li$^+$, TFSI$^-$ and SN molecules in the PEO matrix. Such low density was determined to ensure that the PEO melt does not suffer a potential energy trap arising from contact with other molecules. The number of particles in each loaded system is shown in Table **2**.

**Table 2.** Number of atoms for each loaded electrolyte system in our simulations

| wt% of SN | Compositional Details | | | | Total number of atoms |
|---|---|---|---|---|---|
| | # PEO Chains | # Li | # TFSI molecules | # SN molecules | |
| 0 | | | | 0 | 41600 |
| 2.2 | | | | 100 | 42600 |
| 10 | 100 | 400 | 400 | 465 | 46250 |
| 15 | | | | 740 | 49000 |
| 20 | | | | 1050 | 52100 |

**2.3. Equilibration Protocol.** The complete SN-PEO-LiTFSI electrolyte systems containing appropriate proportions of polymers, ions, and plasticizers at each loading of SN were subjected to an equilibration procedure described previously.[32,50] The initially built systems were minimized using the steepest descent method with a step size of 0.01 nm. The minimization was treated as converged when the maximum force was smaller than 10 kJ mol$^{-1}$ nm$^{-1}$. The minimized structures were then subjected to a short NVT ensemble for 50 ps at 425 K using Berendsen thermostat[62] with a coupling constant of 1 ps. Further, a 10 ns NPT simulation was carried to bring out the low dense system to an experimentally comparable dense state point. During this NPT stage, the system was coupled to a thermostat using the Nosé-Hoover algorithm[63–65] with a damping relaxation time of 1 ps and Parrinello-Rahman barostat[66] with a damping relaxation time of 2 ps. The temperature was annealed to required values for systems having different wt%s of SN. The above-described protocol ensured appropriate equilibration of the electrolyte systems at the desired state point before long production runs of lengths 250 ns. The equations of motion were integrated using the leapfrog algorithm, and the trajectories were saved every 1 ps. We performed ten parallel runs with a saving frequency of 10 fs to analyze dihedral autocorrelation functions. Figure **1** displays a snapshot of equilibrated SPE of EO:Li = 12.5:1 and 15 wt % of SN containing PEO chains, Li$^+$, TFSI$^-$ and SN molecules.

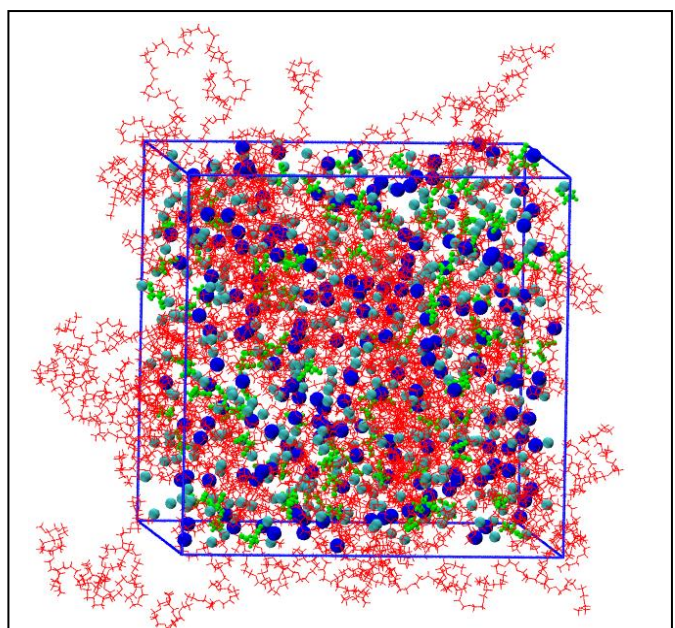

**Figure 1**: Snapshot of an equilibrated SN-PEO-LiTFSI at 425 K having EO:Li = 12.5:1 and 15 wt% of SN. The PEO chains are shown in red as line representation, N atom of TFSI$^-$ ions in cyan VDW representation, SN molecules in green CPK representation, and Li$^+$ ions in blue VDW representation.

The cut-off for LJ interactions was taken as 1.2 nm, and the electrostatic interactions were calculated using the particle mesh Ewald (PME) method.[67] Periodic boundary conditions were applied in all three directions. A summary of the simulation setup and equilibration densities is provided in Table **3**. The densities obtained from the simulations are in good agreement with experimental densities. Furthermore, a slight change in density is observed when adding SN with different loadings than SN-free electrolytes.

**Table 3**. A summary of the simulation setup and average density in g/cm$^3$. The values shown in parenthesis are the standard deviation on the average density.

| EO: Li | T (K) | 0 wt% | 2.2 wt% | 10 wt% | 15 wt% | 20 wt% |
|---|---|---|---|---|---|---|
| 12.5:1 | 375 | 1.294(4) | 1.282(4) | - | 1.215(4) | - |
| | 400 | 1.271(4) | 1.259(4) | - | - | - |
| | 425 | 1.249(4) | 1.236(4) | 1.195(4) | 1.168(4) | 1.143(4) |
| | 450 | 1.227(4) | 1.214(4) | - | - | - |
| | 475 | 1.206(4) | 1.193(5) | - | 1.123(5) | - |
| | 500 | 1.184(5) | 1.171(5) | - | - | - |
| | 525 | 1.163(5) | 1.150(5) | - | 1.078(5) | - |
| | 550 | 1.143(5) | 1.129(5) | - | - | - |
| | 575 | 1.122(6) | 1.108(6) | - | 1.033(6) | - |

## 3. RESULTS AND DISCUSSION

**3.1. Mean-Squared Displacement and Diffusion Coefficient.** Following the equilibration protocol described above, we performed the molecular dynamics simulations at 1 bar pressure for temperatures ranging from 375 K to 575 K at different loading of 0, 2.2, and 15 wt%s of SN molecules. Moreover, we also performed MD simulations at 425 K with various loadings of SN in PEO-LiTFSI electrolyte to understand the loading effect of SN on ion transport. The transport characteristics of Li and TFSI ions were then studied by calculating the mean-squared displacement using $MSD(t) = \left\langle \left(\vec{R}(t) - \vec{R}(0)\right)^2 \right\rangle$ where $\vec{R}(t)$ denotes the position vector of ions at a specific time $t$ and $\langle ... \rangle$ represents the ensemble average over number of particles and all possible time origins. The MSDs for different electrolyte systems at different wt%s and temperatures are shown in Figure **2**.

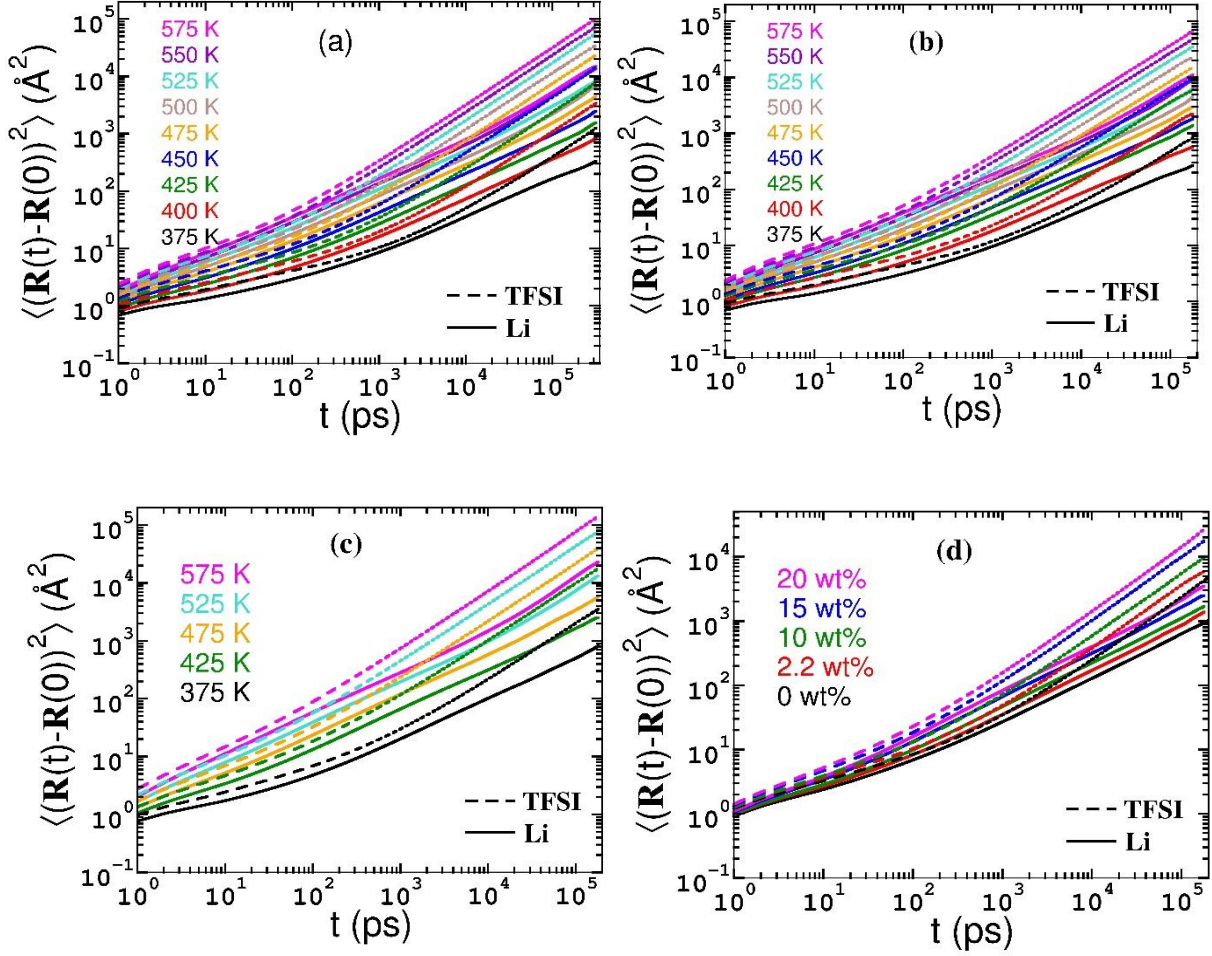

**Figure 2**: The mean squared displacement of Li and TFSI ions at different temperatures for (a) 0 wt%, (b) 2.2 wt%, (c) 15 wt% loading of SN, and (d) for different loading of SN at 425 K. The dashed and continuous curves represent the TFSI and Li-ions, respectively.

The self-diffusion coefficient was then calculated using the Einstein's relation,

$$D = \lim_{t \to \infty} D^{\text{app}}(t) = \lim_{t \to \infty} \frac{\left\langle \left(\vec{R}(t) - \vec{R}(0)\right)^2 \right\rangle}{6t}, \_\_\_\_\_\_(2)$$

where, $D^{\text{app}}(t)$ is the time-dependent apparent diffusion coefficient. The MSDs were calculated from long enough trajectories to reach the diffusive regimes. The exponents, $\beta$ as in $\left\langle \left(\vec{R}(t) - \vec{R}(0)\right)^2 \right\rangle \sim t^\beta$ are shown in Figure **S2**. The Einstein relation holds true only for longer timescales in the diffusive regime at timescales ideally as $t \to \infty$. We calculated the standard deviation on the diffusion coefficient by dividing the diffusive part of the MSD curve into several blocks of equal sizes.

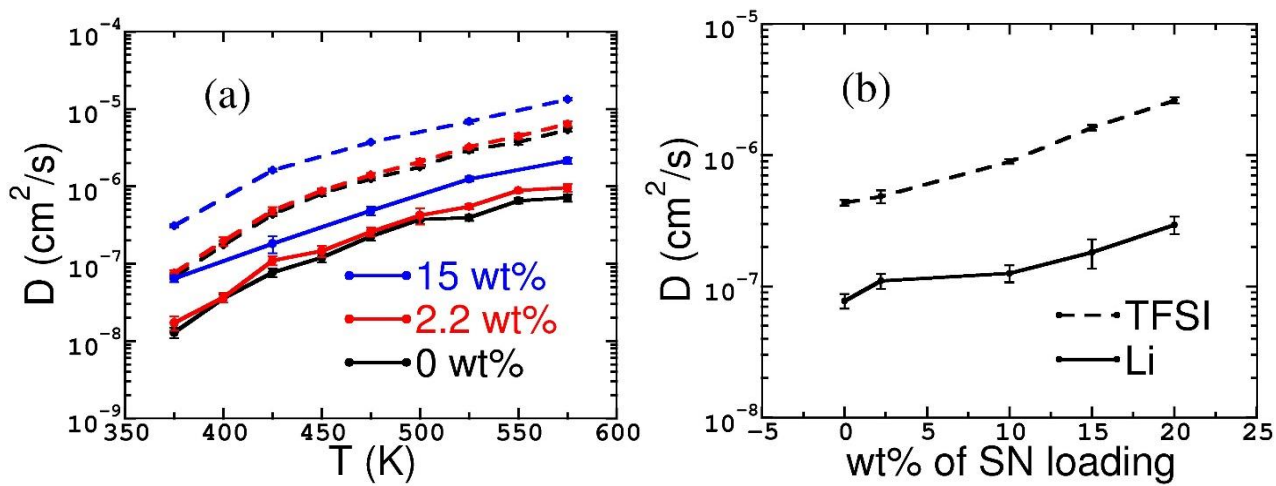

**Figure 3**. (a) The diffusion coefficient of $Li^+$ and $TFSI^-$ ions at 0, 2.2, and 15 wt%s of SN loadings as a function of temperature, and (b) Diffusion coefficient of $Li^+$ and $TFSI^-$ ions at 425 K temperature as a function of weight percentage of SN in PEO-LiTFSI electrolytes. The error bar on many data points is too small to be visible, and the lines guide the eye.

From the MSD results presented in Figure **2**, it is observed that the motion of the TFSI ion transforms from subdiffusive to the diffusive region after $10^4$ ps. However, the Li-ion remains subdiffusive for a relatively longer timescale compared to the diffusive behavior of TFSI ions. Further, the onset of diffusive regimes arrives at a much shorter timescale at higher temperatures. Also, with an increase in temperature, both ions move readily towards the diffusive region. Interestingly, we observed that the MSDs for both cation and anion are significantly higher and increased with SN loading. Explicitly, we observed that adding SN leads to a faster transformation of the subdiffusive region to a long-time linear diffusive part.

To calculate the self-diffusion coefficient correctly from the MD simulations, we used the long-time MSDs. The self-diffusion coefficients of Li and TFSI ions as a function of the temperature at 0, 2.2, and 15 wt%s of SN loadings and as a function of wt%s of SN at 425 K were shown in Figure **3**. It follows that the diffusion coefficient increases exponentially with temperature, consistent with the trends observed for MSD curves. Quantitatively, the value $D$ for Li and TFSI ions was found to be $7.7 \times 10^{-8}$ and $4.3 \times 10^{-7}$ cm$^2$/s in SN free SPE (i.e., pure PEO melt) which increases to $2.2 \times 10^{-7}$ and $1.4 \times 10^{-6}$ cm$^2$/s, respectively when loaded with 20 wt% of SN. It is observed that the diffusion coefficient is found to be higher for TFSI ions than that of Li-ions due to the difference in the size of ions. The TFSI ions have a larger size than Li-ions, which results in a sparsely distributed charge on the TFSI ions compared to that of Li-ions. Consequently, the TFSI ions interact less with their counterions and neighboring positive charge clusters. Therefore, the diffusivity of TFSI is naturally higher than Li-ions. Overall, we observed an increase in ion mobility with increased temperature and SN loading.

Qualitatively, the Nernst-Einstein relation suggests that if the ions in SPE are completely uncorrelated, the ionic conductivity of polymer electrolyte increases correspondingly with diffusion coefficients of anion and cation. However, we did not undertake ionic conductivity calculations explicitly, which is beyond the scope of current work. In the next section, we explain the reasons behind the change in diffusivities and understand ion transport mechanisms.

## 3.2. MECHANISMS UNDERLYING ION MOBILITIES

A variety of transport mechanisms are studied both experimentally and theoretically for the ion diffusion in polymer electrolytes, such as analyzing the interaction of Li-ion with the counterions and the EO unit of polymer segment by ion hopping.[14,18,32,68,69] To investigate the reason behind the relaxation phenomena contributing to the ion diffusivity, we studied the ion-pair relaxation by ion-pair autocorrelation function and polymer dynamics by dihedral autocorrelation function. The PEO host's local relaxation and segmental motion are needed to efficiently transport the ions in polymer electrolytes.[70] Despite many phenomena taking part in ion transport, the ion-pair correlation, and polymer segmental mobility were identified as major contributing factors in influencing ion transport.[13,14,16,18,32–35,71]

### 3.2.1. The Li-TFSI Ion-pair Autocorrelation Functions.

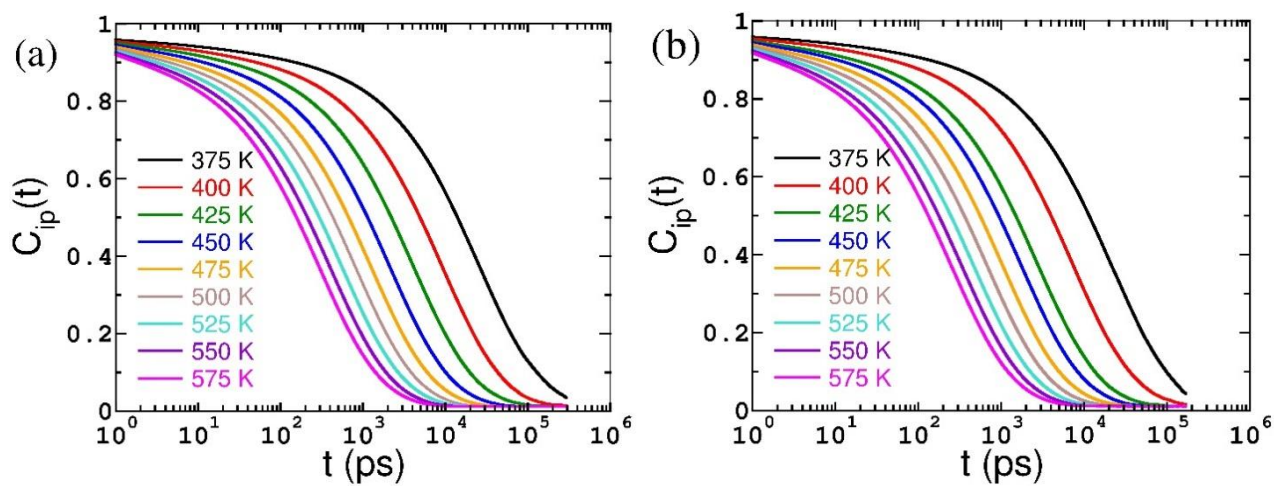

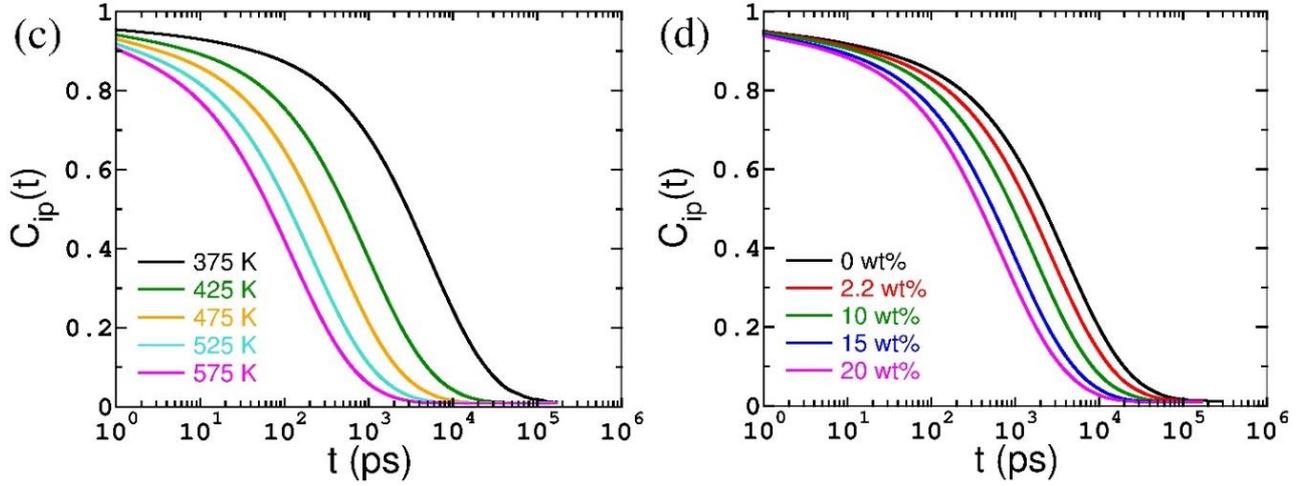

**Figure 4**: The ion-pair autocorrelation functions of Li-TFSI ion-pairs for (a) 0, (b) 2.2, and (c) 15 wt%s of SN loadings at different temperatures, respectively, and (d) The ion-pair autocorrelation functions of Li-TFSI ion-pairs for 425 K at different wt%s of SN loadings.

The primary role of SN in the SPE was to influence the diffusivity of ions, as discussed in the previous section. A higher value of diffusion coefficient of ions is expected to result in a significant increase in ionic conductivities. However, the diffusivities and ionic conductivities ultimately depend on cation-anion correlations and the underlying relaxation phenomena. To gain a fundamental understanding of the above, we begin our analysis by asking, "what is the role of ion-pair correlation timescales on ionic diffusivities?" and examine the relationship between the ion-pair relaxation timescales and the ion transport. In simulations, by constructing a population variable to label whether or not a tagged ion-pair is intact, we calculated the ion-pair autocorrelation function $C_{ip}(t)$ as,

$$C_{ip}(t) = \frac{\langle h(t)h(0)\rangle}{\langle h(0)h(0)\rangle}, \quad\text{———(5)}$$

where, the population variable $h(t)$ is assigned a value 1 when an ion-pair is found intact (i.e., TFSI$^-$ has a Li$^+$ ion in its first coordination shell of 10 Å based on the g(r) shown in Figure **S3**, and vice-versa), and 0 otherwise. The angular brackets account for the ensemble averaging of time correlation function over all possible ion-pairs and time origins. To quantify the ion-pair relaxation timescales, we fitted the simulated results for $C_{ip}(t)$ to the Kohlrausch-Williams-Watts (KWW) stretched exponential function of the form $C_{ip}(t) = e^{-\left(\frac{t}{t_{KWW}}\right)^{\beta_{KWW}^{ip}}}$, where $t_{KWW}$ and $\beta_{KWW}^{ip}$ are the fitting parameters.

The $C_{ip}(t)$ basically quantifies the probability that the two pairs Li-TFSI are intact at time $t$, given that it was intact at time zero. This is calculated from simulated trajectories by observing the joint occurrence of two non-zero populations separated by time $t$. Therefore, at equilibrium, the probability of having a specific pair bonded in a large system is zero, i.e., $C_{ip}(t) = 0$.[75] Here, we took a cut-off distance of 10 Å between the Li and TFSI ions based on the extent of the first coordination shell. We note that while the choice of cutoff has a strong dependency on the absolute values of ion-pair relaxation times, the qualitative features of $\tau_{ip}$ as a function of temperature or loading are generally unaffected as long as the cutoff represents the first coordination shell correctly (see Figure **S4**). By construction, $C_{ip}(0) = 1$ to indicate that ion-pairs are intact at the initial time. The $C_{ip}(t)$ between the pairs Li-TFSI is shown in Figure **4** for different wt%s of SN at 425 K and the different temperatures at 0, 2.2, and 15 wt%s of SN loadings. The $C_{ip}(t)$ decays rather slowly reaching approximately 0.90 - 0.95 within 1 ps and to zero on the scale of $10^5$ ps. We observed similar behavior in both cases of temperature variation and wt% variation. Explicitly, we observed that $C_{ip}(t)$ decays rapidly when polymer electrolyte is loaded with SN as compared to pure PEO melt or the temperature, which means that ion aggregation of Li-TFSI pairs is depressed significantly, indicating that Li-ions are interacting only for a shorter time with TFSI ions.

The ion-pair relaxation time $\tau_{ip}$ was calculated as the total area under the curve $C_{ip}(t)$ such that,

$$\tau_{ip} = \int_0^\infty C_{ip}(t)\,dt = \int_0^\infty \exp\left[-\left(\frac{t}{t_{KWW}}\right)^{\beta_{KWW}^{ip}}\right]dt = t_{KWW}\Gamma\left(1 + \frac{1}{\beta_{KWW}^{ip}}\right), \_\_\_\_\_(6)$$

where $\Gamma$ denotes the Gamma function.

The results of $\tau_{ip}$ at different SN loadings and wt%s are displayed in Figure **S5**. We notice that the ion-pair relaxation time decreased monotonically with an increase in SN loading and temperature. This indicates faster ion-pair relaxations consistent with the $C_{ip}(t)$. The rate of change of $\tau_{ip}$ with temperature is similar to the corresponding changes in the diffusion coefficient as shown in the Figure **S6**.

To gain deeper insights into the role of ion-pair relaxation timescales on the diffusivities, we have plotted the diffusion coefficient as a function of $\tau_{ip}$ as shown in Figure **5**. The diffusion coefficient is found to decrease with the ion-pair relaxation timescales consistently for all the loadings of SN in the electrolyte. The slower relaxations hinder the motion of both the ions in the electrolyte because the tendency of ion-pairing is more pronounced. Therefore, we observed a higher degree of correlations between the diffusion coefficient of Li and TFSI ions and ion-pair relaxation timescales. Further, to quantify such a behavior, we have analyzed the degree of correlations by fitting the diffusion coefficient of ionic species to $D = \alpha_{ip}/\tau_{ip}^{\gamma_{ip}}$ where $\gamma_{ip}$ is ion-pair relaxation exponent.

For the lithium and TFSI ions, we find that the degree of correlation is 0.85 and 0.91, respectively. These exponents point out slight deviations from the ideal Stoke-Einstein (SE) behavior that corresponds to an exponent 1 shown as a dashed line in Figure **5**. We note that in line with the unchanged qualitative features of $\tau_{ip}$ with the temperature at different cutoffs, the trends observed for the exponent $\gamma$ are also insensitive to the cutoff chosen in the first coordination shell (see Figure **S4**), consistent with the previous studies.[76] When fitted with the ideal SE relation, the corresponding parameter is given as $\alpha = 0.00058$ for Li-ions and $\alpha = 0.0036$ for TFSI ions. The degree of deviation from the ideal Stoke-Einstein behavior is somewhat higher for the Li-ions than the TFSI ions, as noticed in Figure **5(a)** and **5(b)**. The same degree of uncorrelation is quantified by plotting $D\tau_{ip}$ vs $\tau_{ip}$ in Figure **6**, where the dashed line indicates the ideal Stokes-Einstein relation and the continuous line was drawn to guide the eye. The degree of uncorrelation is indeed small for TFSI but very high for Li-ions.

Since Li and TFSI ions form ion-pairs, the corresponding relaxation phenomenon is expected to influence equally the diffusion behavior of both the ion types. However, the deviation is higher for the Li-ions and suggests that Li-ions are probably influenced by a different transport mechanism likely caused by the PEO polymer chains. Further, the results presented in Figures **4** and **5** conclusively point out the underlying deviations from the ideal SE relation, which also confirms that the ion-pair relaxation is not yet sufficient for explaining the diffusion coefficient result, which is even worse for the Li case.

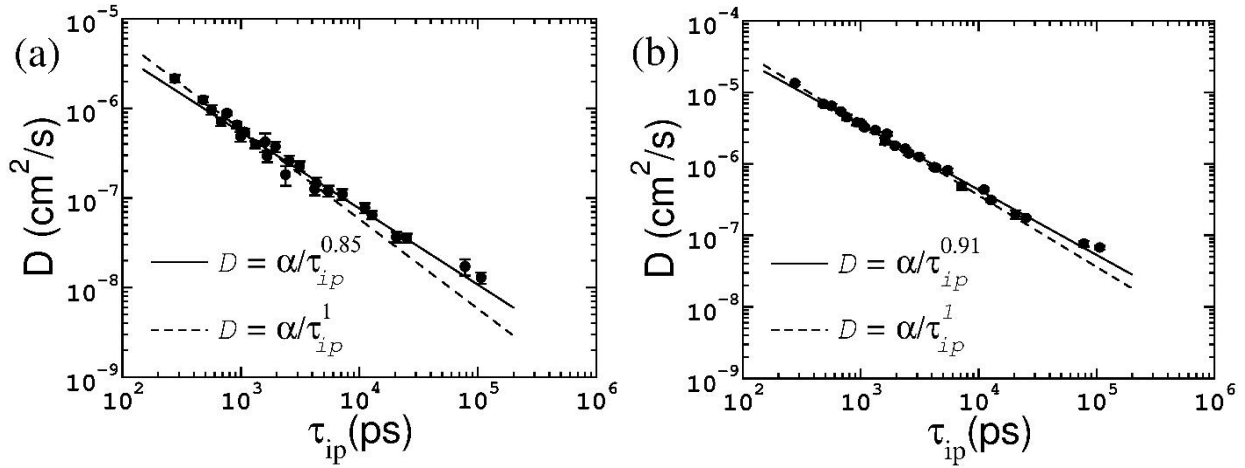

**Figure 5**: The diffusion coefficient as a function of the ion-pair relaxation time for (a) Li, and (b) TFSI ions obtained at all temperatures and loadings of SN studied in this paper. The solid lines represent the fitting of the simulation data to eq $D = \alpha/\tau_{ip}^{\gamma_{ip}}$, where $\alpha$ and $\gamma$ are fitting parameters. The obtained fitting exponent is 1 for Li-ion indicating agreement with the ideal SE behavior and filled symbols represent the simulation data. The fitted $\alpha$ values are: $\alpha = 0.00058$ and $0.0036$ for Li and TFSI, respectively, when the exponent is fixed at a value of 1.

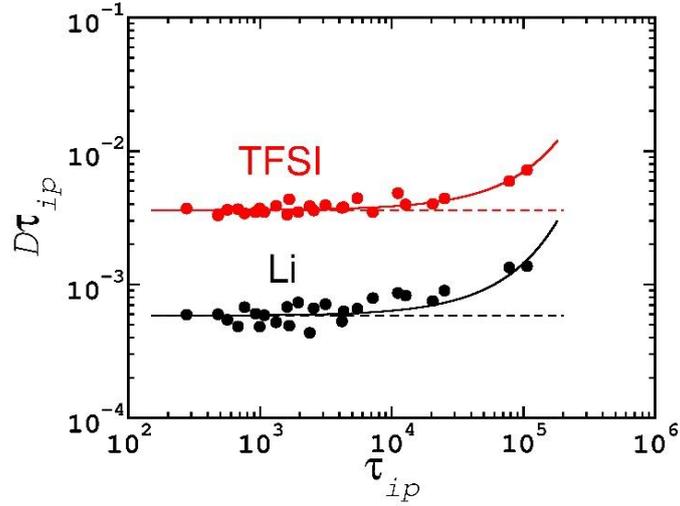

**Figure 6**: Verification of the deviation from SE relation: comparison of $D\tau_{ip}$ vs $\tau_{ip}$ for simulation data of Li and TFSI ions at all temperatures and wt%s. The dashed lines represent ideal SE behavior, and solid lines are drawn to guide the eye using equations $D\tau_{ip} = 5.8 \times 10^{-4} e^{9.1 \times 10^{-6} \tau_{ip}}$ and $D\tau_{ip} = 3.6 \times 10^{-3} e^{6.6 \times 10^{-6} \tau_{ip}}$, respectively for Li and TFSI ions.

To understand how ionic diffusivities are correlated to the ion-pair relaxation phenomenon when the loading of SN is changed, we further decomposed the data as shown in Figure **6**. One of the main questions in this regard is: How do the correlations between diffusivities and relaxation timescales change at different loadings when changing the temperature in the polymer electrolyte?. Further, we seek to understand how differently does the ion-pair relaxation phenomenon influence the ion transport behavior when changing the temperature at a particular wt% (Figure **7**(a-c)) versus when changing the loading of SN at a particular temperature (Figures **7**(d)) in the polymer electrolyte. Not surprisingly, we observed that the decomposed data confirm the same physics that the deviations from ideal SE relation are more significant in the case of Li-ions but relatively lower in TFSI ions, consistent with the results presented in Figure **6**. Interestingly, we observe that the correlations approach towards a one-to-one correspondence scenario at higher loadings of SN for both types of ions indicates smaller deviations from the ideal SE relation at higher loadings. However, the correlations worsen for lithium ions but better for TFSI ions when changing the loading of SN in the polymer electrolyte at 425 K (Figure **7**(d)).

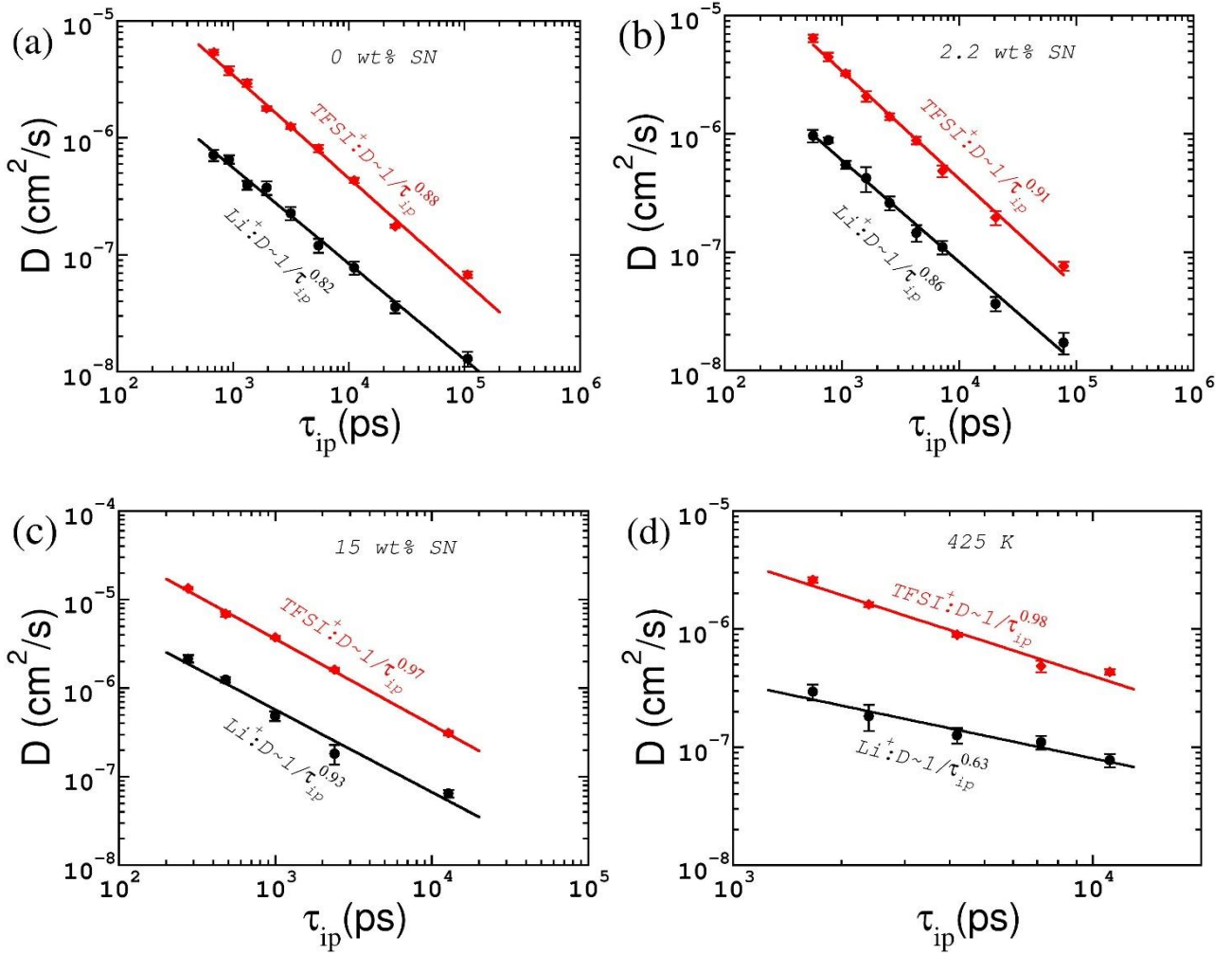

**Figure 7**: The diffusion coefficient as a function of the ion-pair relaxation time for SN loadings of (a) 0, (b) 2.2, and (c) 15 wt%s at different temperatures, and (d) for different loadings at 425 K.

Together, the results presented in Figures **5-7** collectively suggest that the ion diffusivity of both types of ions is only *either partially or poorly* described by the underlying ion-pair relaxation phenomenon in most of the scenarios and *entirely* in only two scenarios for the TFSI ions: (i) Temperature-induced faster diffusion at a loading of 15 wt% with an exponent of 0.97, and (ii) SN loading induced faster diffusion at 425 K with an exponent of 0.98. The *partial or poor* correlations between ion transport and underlying ion-pair relaxation phenomenon in the case of lithium ions indicate that the diffusivity of Li-ions may only be described by invoking different other transport mechanisms. The above analysis suggests that no polymer groups interact with TFSI ions favorably to provide ion transport pathways, but lithium may have favorable transport pathways along the backbone of polymer segments. Therefore, we further explored the dihedral angle autocorrelation functions because the Li-ion is expected to interact more closely with the polymer chains than TFSI ions which are only interacting through ion-pair interaction with Li-ions. We, therefore, examine the hypothesis that the interplay of Li-ion interactions with polymer segments is possibly responsible for Li-ion transport in the following section.

## 3.2.2. Polymer Segmental Motion Through Dihedral Autocorrelation Functions.

As discussed in the previous section, the polymer segmental motion of PEO chains is also crucial in understanding ion mobility. Moreover, from the results of the diffusion coefficient, it appears that the effect of SN loading has the same qualitative effect as by the temperature. Since the temperature affects the polymer segmental dynamics, loading of the SN may also affect the polymer dynamics. Therefore, we have examined the polymer dynamics and their correlation with ion mobility/diffusivity.

To understand this behavior, we calculated the dihedral autocorrelation function $C_\phi(t)$ involving C-O-C-C atoms in PEO chains as[32]

$$C_\phi(t) = \frac{\langle \cos\phi(t)\cos\phi(0)\rangle - \langle \cos\phi(0)\rangle^2}{\langle \cos\phi(0)\cos\phi(0)\rangle - \langle \cos\phi(0)\rangle^2}, \quad\text{———(7)}$$

where $\phi(t)$ is the dihedral angle of C-O-C-C atoms in the PEO chain at time $t$ and $\langle ... \rangle$ represents the ensemble average over all possible C-O-C-C dihedral angles and time origins. By definition, the $C_\phi(0) = 1$ and decays to zero much faster than the ion-pair time correlation function $C_{ip}(t)$. Due to the inherent faster polymer dynamics, the $C_\phi(t)$ rapidly decays to almost 0.6-0.8 within just 1 ps, depending on the temperature. Therefore, we saved the trajectories with a frequency of 10 fs to capture the fast relaxations. We considered 10 such independent trajectories of high saving frequency, and the resulting average $C_\phi(t)$ was calculated for lag times up to 1 ps. Beyond 1 ps, the $C_\phi(t)$ was calculated using a single long trajectory of 250 ns for a given loading and temperature.

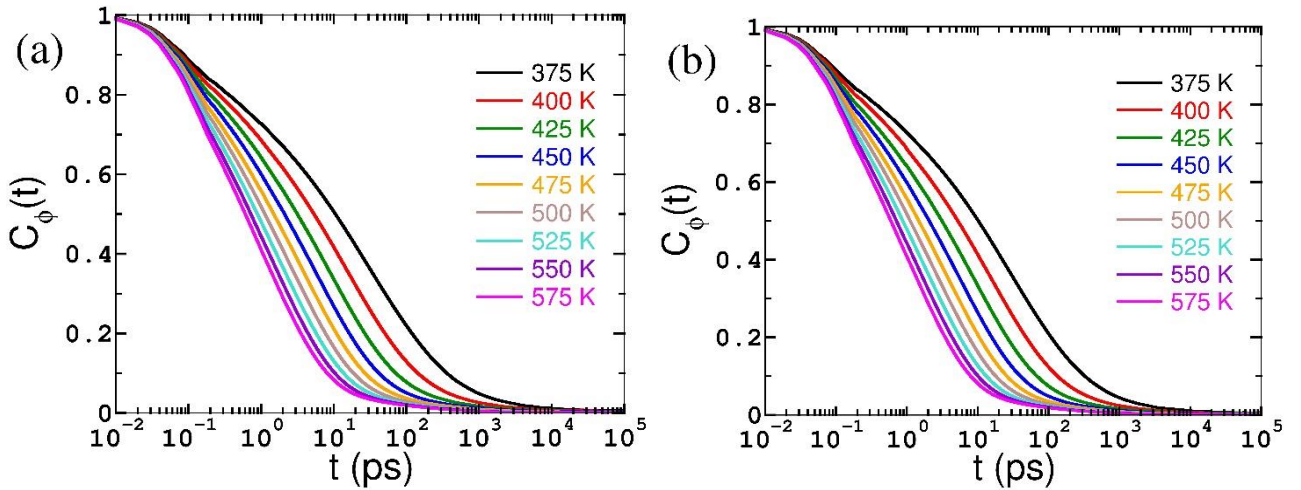

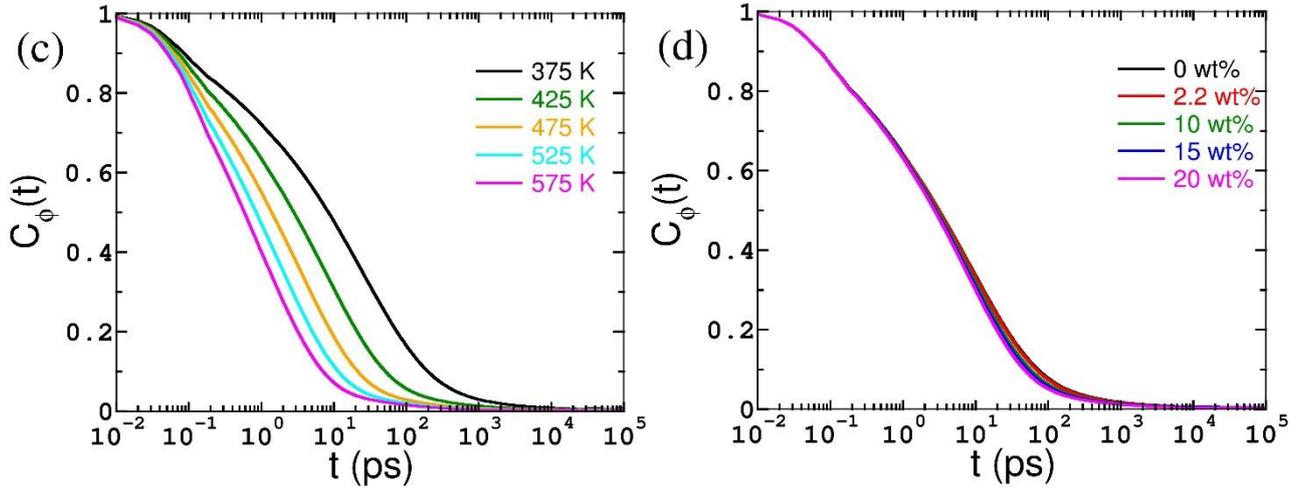

**Figure 8**. The dihedral angle autocorrelation functions of PEO chains for SN loadings of (a) 0, (b) 2.2, and (c) 15 wt%s at different temperatures, and (d) dihedral autocorrelation function for different loadings of SN molecules at 425 K.

The results of $C_\phi(t)$ for different temperatures and loading are shown in Figure **8**. We observed the dihedral autocorrelation function $C_\phi(t)$ to change significantly with temperature for a given loading. However, the changes in $C_\phi(t)$ are minimal with respect to the loading of SN at 425 K. We observed that at high temperature, there is a sharp decay in $C_\phi(t)$ whereas, with an increase in loading of SN, the $C_\phi(t)$ does not decay significantly as compared to pure PEO melt. The $C_\phi(t)$ was found to decrease consistently with increasing temperatures. The $\tau_\phi$ changes exponentially but a more rapid exponential decay was observed as a function of temperature than wt%. To understand the effect of polymer segmental relaxation on the diffusion of ions, we plotted the corresponding diffusion coefficient of ions with the dihedral angle relaxation time $\tau_\phi$ as shown in Figure **9**.

Similar to ion pair autocorrelation function, here, we have fitted the $C_\phi(t)$ with KWW stretched exponential functions and calculated the dihedral relaxation times as, $\tau_\phi = t_{\text{KWW}} \Gamma\left(1 + \frac{1}{\beta_{\text{KWW}}^\phi}\right)$. We observed $\tau_\phi$ decreases exponentially (shown in Figure **S7** in SI) with the loading of SN in the system, which indicates the presence of SN makes the polymer more dynamic, which indirectly is expected to help in faster transport of ions. Similarly, $\tau_\phi$ also decreases significantly with the increase in temperature. Qualitatively, the behavior is like an Arrhenius, indicating that there could be correlations between the diffusion coefficient and dihedral angle relaxation timescales. To quantify the degree of correlation between the $D$ and dihedral angle relaxation time, we fitted $D$ vs $\tau_\phi$ with the equation $D = \alpha/\tau_\phi^{\gamma_\phi}$ and obtained the fitting parameter α and exponent $\gamma_\phi$ as shown in Figures **9**(a-b) for all wt%s and temperatures.

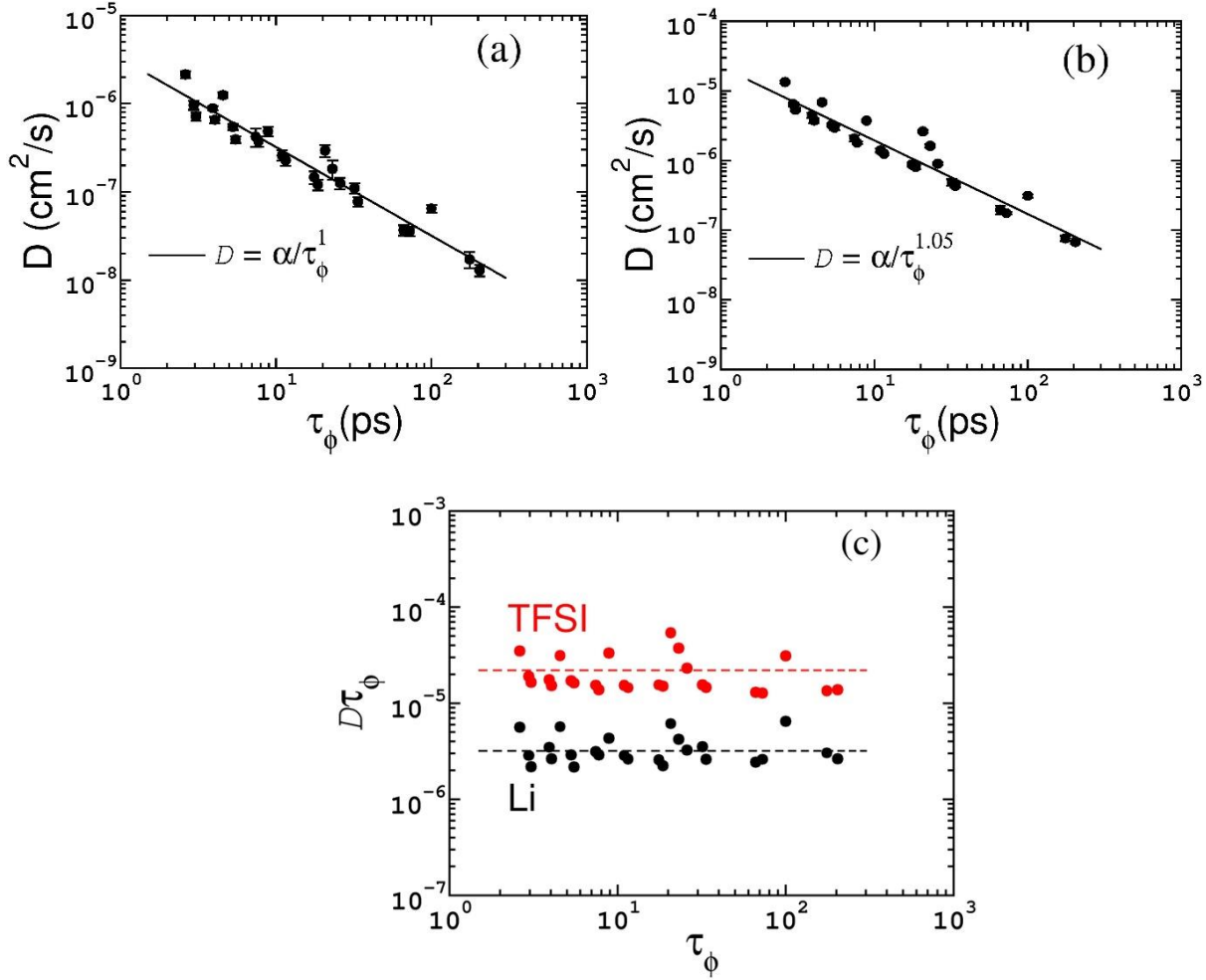

**Figure 9**: The diffusion coefficient as a function of the dihedral angle relaxation time for (a) Li, and (b) TFSI ions. The solid lines represent the fitting of the simulation data to eq $D = \alpha/\tau_{ip}^{\gamma_{ip}}$, where $\alpha$ and $\gamma_{ip}$ are fitting parameters. The obtained fitting exponent is 1 for Li-ion indicating agreement with the ideal SE behavior and dashed lines represent fitting of the simulation data to ideal SE behavior for TFSI ion with eq $D = \alpha/\tau_{ip}$, where $\alpha$ is a fitting parameter. The fitted $\alpha$ values are: $\alpha = 0.0000032$ and 0.000022 for Li and TFSI, respectively, when the exponent is fixed at a value of 1. (c) Verification of the deviation from SE relation: comparison of $D\tau_\phi$ vs $\tau_\phi$ for simulation data of Li and TFSI ions at all temperatures and wt%s. The dashed lines represent ideal SE behavior, and solid lines are drawn to guide the eye using equations $D\tau_\phi = 0.32 \times 10^{-5}\, e^{3.99\times 10^{-4}\tau_\phi}$ and $D\tau_\phi = 2.2 \times 10^{-5}\, e^{-1.7\times 10^{-3}\tau_\phi}$, respectively for Li and TFSI ions.

Interestingly, we observed that the diffusion coefficient of Li correlates excellently with dihedral angle relaxation timescales, but for TFSI ions, a slight deviation is observed with an exponent close to 1.0 as shown in Figures **9**(a-b). This can be understood by invoking the fact that the lithium ions form a strong coordination shell with ether oxygens of polymer (see Figure **S3**), which helps them hop along the polymer backbone. Such a transport mechanism of lithium ions is related to the dihedral angle relaxation timescales that capture the polymer segmental motion. However, the TFSI ions do not form a coordination shell with ether oxygens of polymer but only with the lithium ions. With the addition of SN, the polymer dynamics is changing more strongly in comparison to the ion-pair relaxation timescales. As a consequence, the effect of SN is more on the transport of Li ions compared to that of TFSI ions. Therefore, the diffusion coefficient of lithium ions is correlated more closely to the polymer dynamics and the diffusion of TFSI ions is correlated more closely to the ion-pair relaxation timescales.

If we compare Figure **6** with Figure **9**(c), the ion-pair relaxation timescales deviate from the SE relation more than the dihedral angle relaxation timescales. The differences observed between Figures **6** and **9**(c) raise the following question: "Is polymer dynamics more important that the ion-pair relaxation timescales?" We recall that while the $\tau_{ip}$ based mechanisms gave us deeper insights on whether or not there can be deviations from $D$ vs $\tau$ relation, the $\tau_\phi$ based transport mechanisms did not provide us such insights. These observations apparently signify that the polymer dynamics is likely more important than the ion-pair relaxation timescales but further analysis is required to establish such a hypothesis. One possible approach is to define an effective timescale which is a function of $\tau_{ip}$ and $\tau_\phi$, and analyze the correlations between the diffusion coefficient of ions with the effective timescale. However, we did not carry out such an analysis in this paper that may be considered for a future investigation.

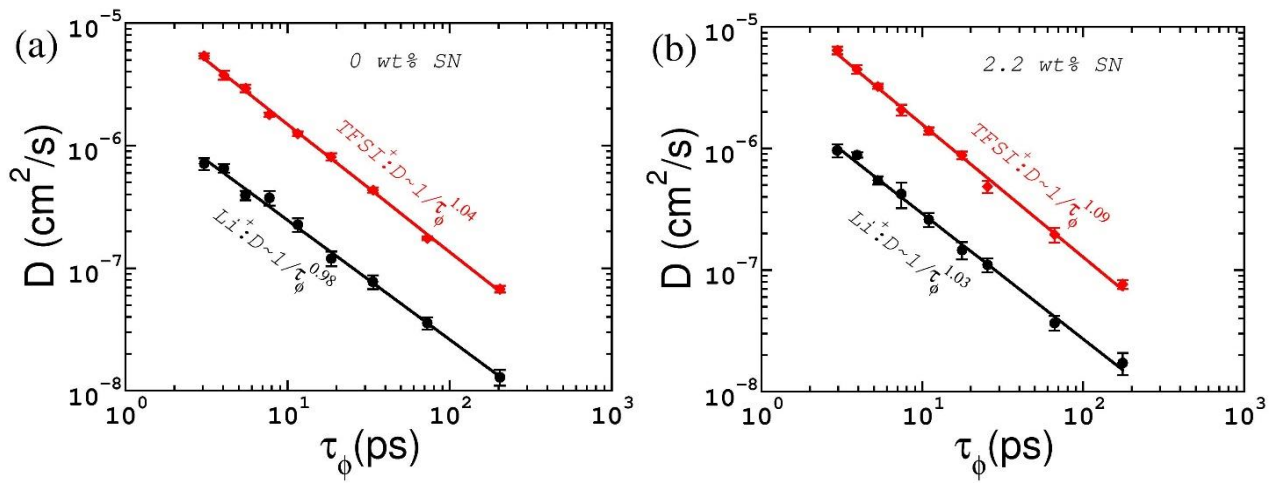

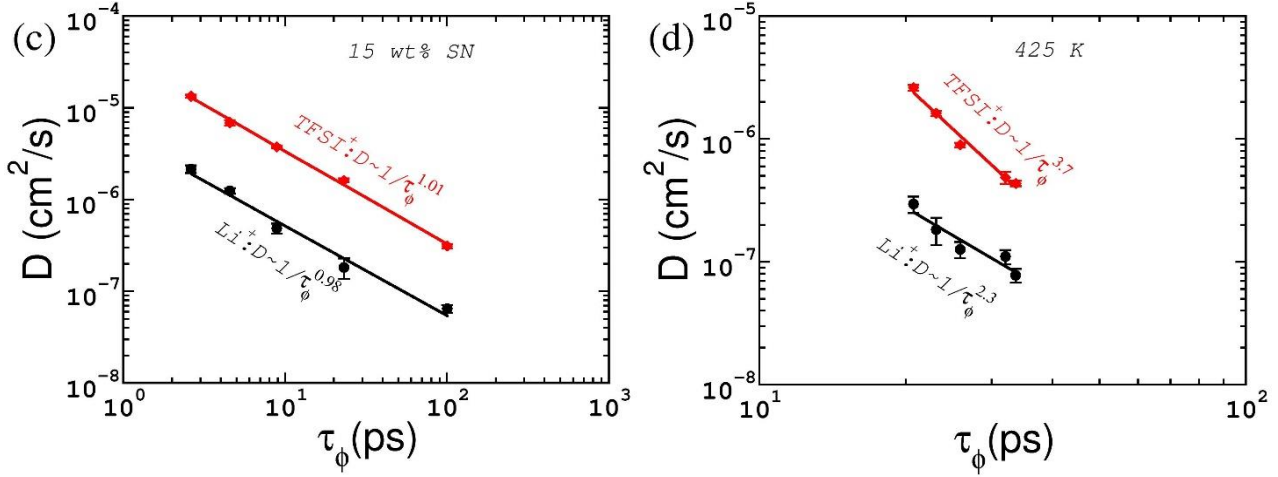

**Figure 10**: The diffusion coefficient as a function of the dihedral angle relaxation time for SN loadings of (a) 0, (b) 2.2, and (c) 15 wt%s at different temperatures, and (d) for different loadings at 425 K.

Consistent with the $C_\phi(t)$ results presented in Figure **8**, we observed a counterintuitive behavior of $D$ vs $\tau_\phi$ at different loadings of SN that the exponent is much higher than 1 as opposed to the correlations found in Figure **7**. The results presented in Figures **5-7** pointed to an ambiguity in conclusively understanding the Li⁺ mobility based on simply the ion-pair relaxation timescales. Based on this, we hypothesized that the dihedral angle relaxation timescales might resolve the issue and explain the Li⁺ ion mobility. From the analysis of temperature effects at 2.2 wt% of SN for Li⁺ ion, the value of the exponent is observed to increase from 0.82 to 0.98 (see Figure **10**(a)), confirming our hypothesis that Li⁺ ions are transported more efficiently with the segmental motion of PEO chains than ion-pair relaxation mechanism. But for the loading effects, the exponent for Li⁺ comes out to be counterintuitive as it increased from 0.63 to 2.3 (see Figure **10**(d)), which is unexpected. Similar observations are made for the TFSI ions as well as for 2.2 and 15wt%s of SN loadings as shown in Figures **10**(b-c).

The effect of SN is seen indirectly through the decrease in relaxation timescales with the loading. This is because the SN helps in separating the ions from their counterions by interacting more strongly with the TFSI ions. The interaction of SN is expected to be strong with ions due to the fact that Nitrogen in SN has a strong negative charge. However, the competing interactions of polymer EO sites with Li outplay the interactions between SN and Li, and therefore the SN molecules influence strongly the TFSI ions, which is clearly seen in Figure 11. The coordination number of SN molecules around TFSI increases with wt% but not around the Li-ions or polymer chains (See Figure **S8**). This is an indication of higher interaction between SN and TFSI and which helps in separating the ion pairs. Therefore, the mobility of ions increases with the addition of SN.

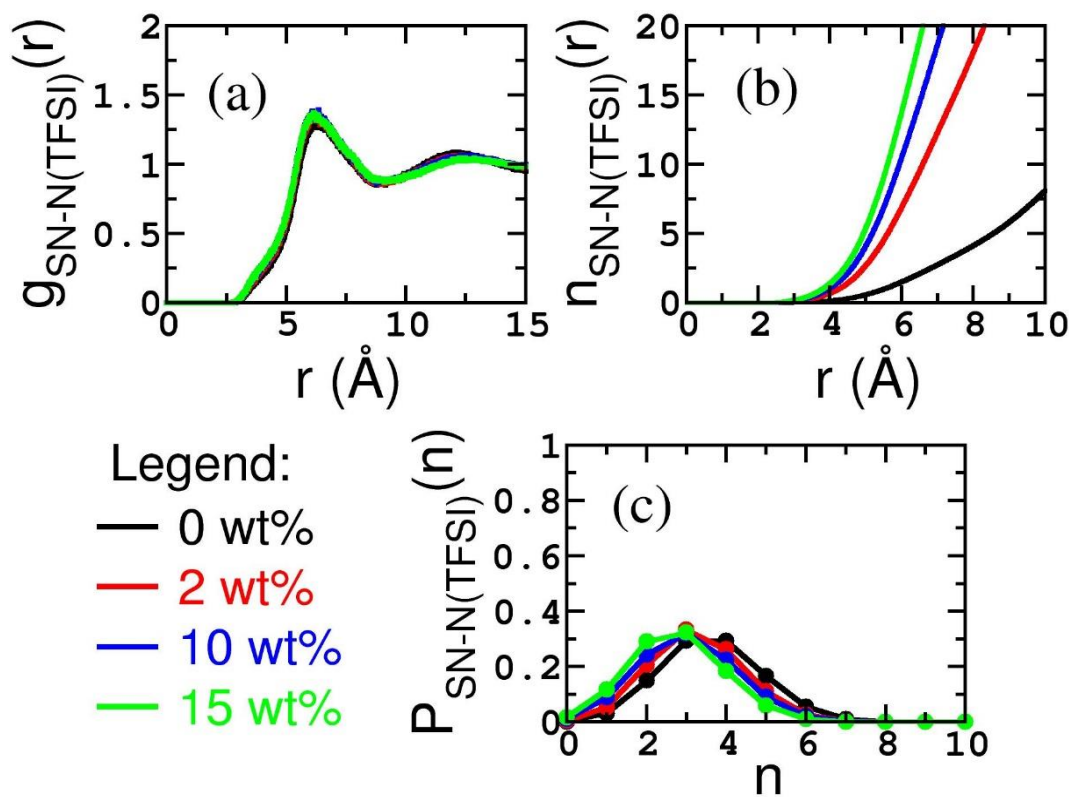

**Figure 11**: (a) Radial distribution function between SN and TFSI, (b) Coordination number for SN around TFSI ions, and (c) Probability, for *i-j* particles, that *n* number of *j*th particles are around *i*th type of particles in the first coordination shell, for SN-TFSI pairs

So, together the direct/indirect influence of SN on polymer relaxation timescales or ion-pair relaxation timescales are not directly helpful in explaining the diffusivity of Li-ions completely. This is a surprising result where the transport mechanism associated with Li-ions is more complex than ion pair relaxation timescales and polymer segmental motion. Further investigation is required to establish this phenomenon completely.

## 4. CONCLUSIONS

To summarize, atomistic molecular dynamics simulations were performed to study the effect of loading of SN on the diffusivity of ionic species in PEO-LiTFSI electrolytes and to understand the underlying transport mechanisms in the solid polymer electrolyte. A sufficient number of SN molecules were dispersed to prepare succinonitrile weight percentages of 0, 2.2, 10, 15, and 20 in the PEO melt along with dissolved LiTFSI salt at a specific salt concentration ratio of EO:Li=12.5:1. The temperature effects were also studied at selected loading of 0, 2.2, and 15 wt%s of SN in the range between 375 – 575 K. To describe the interactions between different molecular species at the atomistic level, we employed the optimized potentials for liquid simulations (OPLS) force field parameters developed by Jorgensen group.[38]

The diffusion coefficient of ionic species was calculated from the long-time diffusive regime of the mean-squared displacements and examined the fundamental origin of ion transport in the SN loaded solid polymer electrolytes by invoking the underlying relaxation phenomena. We find that the diffusivity of both cation and anion increases monotonically with SN's loading and an increase in temperature. The diffusion coefficient of TFSI ions was higher than that of the lithium ions at a given temperature and loading of SN molecules. These results offer an attractive future promise for using succinonitrile as a plasticizer to increase the ionic transport characteristics of PEO-based solid polymer electrolytes.

To explain the mechanisms behind increased diffusion of ionic species, we analyzed different relaxation timescales associated with the Li-TFSI ion-pair dynamics and polymer segmental dynamics by calculating the respective ion-pair dihedral angle autocorrelation functions. We computed the ion-pair and polymer segmental relaxation timescales by fitting the respective autocorrelation functions to Kohlrausch–Williams–Watts (KWW) stretched exponential functions. We observed both $\tau_{ip}$ and $\tau_{\phi}$ to decrease monotonically with an increase in the loading of SN and temperature, revealing accelerated relaxation behavior of SN-PEO-LiTFSI electrolytes. We further investigated the inherent connection between the ion-pair relaxation timescales, polymer segmental timescales, and ionic diffusivities. We found that both the relaxation timescales are necessary to explain the diffusivities of ions completely. While the TFSI ions are found to diffuse in the electrolytes by the influence of primarily the ion-pair interactions, the diffusion mechanism of lithium ions is strongly affected by the interactions of Li-ions coming from both the TFSI ions and the PEO polymer chains. However, the analysis of correlations between diffusion coefficient of ions as a function of wt% of SN loading and different timescales point out unusual SN-induced transport mechanisms in SN-PEO-LiTFSI electrolytes.

Overall, our simulations suggest that SN molecules strongly affect the $Li^+$ ion mobility in poly(ethylene oxide) with dissolved Li-TFSI polymer electrolytes and may boost the choice of PEO-based solid polymer electrolytes possessing higher ionic conductivity for commercial applications. Our deeper analysis led to an interesting question for future investigations that ion transport in SN-loaded electrolytes cannot be described entirely by ion-pairing behavior and polymer segmental motion of PEO chains and that existence of interplay between other distinct transport mechanisms influencing ion transport.

**AUTHOR INFORMATION**

**Corresponding Author**

*E-mail: santosh@iitj.ac.in

**Notes**

The authors declare no competing financial interest.

## SUPPLEMENTARY MATERIAL

See the supplementary material for the detailed information about data supporting the results presented in this paper such as molecular structure, exponent observed in MSD for ions, RDF and coordination number, ion-pair relaxation timescale , and probability of association of ions etc.

## ACKNOWLEDGMENTS

The authors acknowledge the Computer Center of IIT Jodhpur for providing computing resources that have contributed to the research results reported in this paper. S. Mohapatra thanks the Ministry of Human Resource and Development (MHRD) for supporting the work. We acknowledge the financial support of Grant No. SERB CRG/2019/000106. The authors acknowledge stimulating discussions with Prabhat K Jaiswal and Ananya Debnath.